\newcommand{\PLB}[3]{Phys.\ Lett.\ B\ {\bf #1},\ #2 (#3)}

\newcommand{\PRL}[3]{Phys.\ Rev.\ Lett.\ {\bf #1},\ #2 (#3)}
\newcommand{\RMP}[3]{Rev.\ Mod.\ Phys.\ {\bf #1},\ #2 (#3)}

\newcommand{\NAT}[3]{Nature\ {\bf #1},\ #2 (#3)}
\newcommand{\SC}[3]{Science\ {\bf #1},\ #2 (#3)}
\newcommand{\NATPHYS}[3]{Nature Phys.\ {\bf #1},\ #2 (#3)}

\newcommand{\PRA}[3]{Phys.\ Rev.\ A\ {\bf #1},\ #2 (#3)}
\newcommand{\PRB}[3]{Phys.\ Rev.\ B\ {\bf #1},\ #2 (#3)}
\newcommand{\PRE}[3]{Phys.\ Rev.\ E\ {\bf #1},\ #2 (#3)}
\newcommand{\PRC}[3]{Phys.\ Rev.\ C\ {\bf #1},\ #2 (#3)}

\newcommand{\JPB}[3]{J.\ Phys.\ B:\ At.\ Mol.\ Opt.\ Phys.\ {\bf #1},\ #2 (#3)}

\newcommand{\JPC}[3]{J.\ Phys.:\ Cond.\ Mat.\ {\bf #1},\ #2 (#3)}

\newcommand{\EPJB}[3]{Eur.\ Phys.\ J.\ B\ {\bf #1},\ #2 (#3)}

\newcommand{\NJP}[3]{New\ J.\ Phys.\ {\bf #1},\ #2 (#3)}

\newcommand{\diracslash}[1]{#1\llap{/\kern2pt}}

\newcommand{\be}{\begin{equation}}
\newcommand{\ee}{\end{equation}}
\newcommand{\bea}{\begin{eqnarray}}
\newcommand{\eea}{\end{eqnarray}}
\newcommand{\ba}[1]{\begin{array}{#1}}
\newcommand{\ea}{\end{array}}

\documentclass{ws-ijmpb}
\usepackage{graphicx}
\usepackage{verbatim}
\usepackage{amssymb}

\usepackage{color}
\begin{document}
\markboth{Ayan Khan, Saurabh Basu, and B. Tanatar}
{Investigating Dirty Crossover through Fidelity Susceptibility and Density of States}

\title{Investigating Dirty Crossover through Fidelity Susceptibility and Density of States}

\author{Ayan Khan\footnote{ayankhan@fen.bilkent.edu.tr}}
\address{Department of Physics, Bilkent University, Bilkent
06800, Ankara, Turkey}
 \author{Saurabh Basu}
 \address{Department of Physics, Indian Institute of Technology-Guwahati, Guwahati, India}
\author{B. Tanatar}
\address{Department of Physics, Bilkent University, Bilkent
06800, Ankara, Turkey}

\date{\today}
\maketitle
\begin{history}
\received{Day Month Year}
\revised{Day Month Year}
\end{history}

\begin{abstract}

We investigate the BCS-BEC crossover in an ultracold
atomic gas in the presence of disorder. The disorder is incorporated in the 
mean-field formalism through Gaussian fluctuations. We observe evolution
to an asymmetric line-shape of fidelity susceptibility as a function of interaction
coupling with increasing disorder strength 
which may point to an impending quantum phase transition. 
The asymmetric line-shape is further analyzed using the statistical 
tools of skewness and kurtosis. We extend our analysis to density 
of states (DOS) for a better
understanding of the crossover in the disordered environment.
\end{abstract}


\section{Introduction} 
Atomic gases at very low temperature are unique systems where
one can observe the continuous evolution of a fermionic system
(BCS type) to a bosonic system (BEC type) 
by changing the inter-atomic interaction by means of Fano-Feshbach 
resonance \cite{grimm}. The essence of BCS-BEC crossover rests on 
the two-body physics where the  inter-atomic $s$-wave scattering 
length is tuned to influence the interaction from weak to
strong.
The weak coupling limit is characterized by the BCS theory where two 
fermions make a Cooper pair with a long coherence length whereas in 
the strong coupling limit they are closely bound to create composite 
bosons (short coherence length), which can undergo Bose-Einstein 
condensation (BEC). In this evolution process there exists 
a region known as ``unitarity'', where the $s$-wave scattering 
length diverges and the interactions depend only on the inverse 
Fermi momentum ($1/k_{F}$), which causes the physics to become 
independent of any length scale and therefore universal for any 
Fermi system.
The experimental advances to realize this transition has introduced the
possibility of studying the so called BCS-BEC crossover more 
closely from different angles\cite{stringari1,dalibard1,he}.

The situation becomes even more interesting if one assumes the existence 
of random disorder in an otherwise very clean system. In the seminal work 
of Anderson\cite{anderson1}, the localization effect due to disorder was 
predicted for superconductors in strong disorder but the weak disorder 
does not produce any significant effect.
It is immensely difficult to observe the localization or 
the exponentially decaying wave function in electronic systems.
One usually takes the indirect route 
of conductivity measurements to observe the effect of
localization\cite{alek}. 

Ultracold atomic systems with their high level of of controllability 
offer a chance to observe effects of disorder more directly. 
Recently, localization effects have been observed in Bose gases 
($^{87}$Rb and $^{39}$K)\cite{billy1,roati1}. 
Latest experiments are conducted for three-dimensional systems for both 
noninteracting atomic Fermi gas\cite{kondov} of 
$^{40}$K and Bose gas\cite{bouyer} of $^{87}$Rb. 
These experiments have widened the possibility of studying the crossover 
in the presence of disorder\cite{lewenstein1,basu1} experimentally.

Recent studies on unitary Fermi gases in the presence of quenched disorder 
have predicted a paradigm of robust superfluidity in the crossover 
region\cite{orso,sademelo2,sademelo3,khan1,khan} through the 
observation of nonmonotonic condensate fraction\cite{orso,khan1} and 
critical temperature\cite{sademelo3}. 
Further, it was indicated that  
the weak impurity effect can lead to a quantum phase transition (QPT)
by extrapolating the data of critical temperature obtained through 
the study of quantum fluctuations (see Fig.[3] in 
Ref.\,\refcite{sademelo2}). These studies have motivated 
us to look into the BCS-BEC crossover and beyond in the presence 
of disorder. For this purpose we have carried out a systematic study on 
fidelity susceptibility (FS) and density of states (DOS) for the 
dirty crossover.

Recently, behavior of FS, a tool widely 
used in quantum computation, is successfully applied as an 
intrinsic criterion to study QPT and tested
in a variety of models\cite{zanna,You07,Yang07,chen07,gu}. 
Since a QPT is an abrupt change of the ground state 
of a many-body system when a control parameter $\lambda$ of the 
Hamiltonian crosses a critical value $\lambda_c$, it is then quite natural 
to expect\cite{zanna} that the overlap 
$F(\lambda+\delta\lambda,\lambda)\equiv |\langle\Psi(\lambda
+\delta\lambda) | \Psi(\lambda)\rangle|$ between the ground states 
corresponding to two slightly different values of the parameter $\lambda$, 
should manifest an abrupt drop when the small variation $\delta\lambda$ 
crosses $\lambda_c$. Such an overlap, which has been named ``ground-state 
fidelity'' provides a signature of a QPT.
We have shown that this technique can be adequately used in the BCS-BEC 
crossover scenario\cite{khan2}. Although BCS-BEC crossover 
is not a QPT, we observe a smooth nonmonotonic evolution of the fidelity 
susceptibility.

In this article we study the overlap function when the system is subjected 
to a weak random disorder. The atom-atom interaction is modeled by
a short-range (contact) potential. In Ref.~\refcite{khan2} it was 
shown that for the contact
potential the FS is highly symmetric and its full width at half
maximum indicates the crossover boundary. Here we observe a breakdown 
in symmetry (using similar short range inter-atomic interaction) 
with increasing disorder
and we study the asymmetry by means of third and fourth moments 
of the FS distribution. 

It is customary to analyze the electronic, topological and physical 
properties of different materials by a systematic study of the 
density-of-states (DOS). \cite{trivedi1}
We note that the energy gap in DOS for BCS superconductors is same 
as the pairing gap for weak disorder.
The energy gap shows a nonmonotonic behavior 
with increasing disorder strength while the pairing gap is 
gradually depleted.
Here, we calculate the DOS in the BCS-BEC crossover region and obtain 
an interesting behavior at unitarity.
Disorder does not introduce a significant change in DOS at the BCS and BEC 
regimes but a distinct drop in the energy gap at unitarity is observed. 

The rest of this paper is organized as follows. In Sec.~\ref{formalism} 
we give a brief description of the disorder model and sketch of 
the FS calculation. We present our results and provide a general
discussion in Sec.~\ref{result}. We draw our
conclusions in Sec.~\ref{conclusion}. 

\section{Formalism}\label{formalism}
In this section, we provide a sketch of the mean-field theory 
when subjected to static random disorder and later we review the FS calculation in the crossover picture\cite{khan2}.
A more comprehensive description of the disorder formalism can be found in 
Refs.~\refcite{orso,sademelo3,khan1}.

\subsection*{Effect of Disorder in the Mean-field Approximation}
To describe the effect of impurities in a Fermi superfluid evolving from 
BCS to BEC regimes one needs to start from the real space 
Hamiltonian in three-dimensions for an $s$-wave superfluid,
\begin{eqnarray}
 \mathcal{H}(\mathbf{x})&=&\sum_{\sigma}\Phi^{\dagger}_{\sigma}(\mathbf{x})\Big[-\frac{\nabla^2}{2m}-
\mu+\mathcal{V}_{d}(\mathbf{x})\Big]\Phi_{\sigma}(\mathbf{x})\nonumber\\
&+&\int dx'\mathcal{V}(\mathbf{x},\mathbf{x'})\Phi_{\uparrow}^{\dagger}(\mathbf{x'})
\Phi_{\downarrow}^{\dagger}(\mathbf{x})\Phi_{\downarrow}(\mathbf{x})\Phi_{\uparrow}(\mathbf{x'}),\label{H}
\end{eqnarray}
where $\Phi_{\sigma}^{\dagger}(\mathbf{x})$ and 
$\Phi_{\sigma}(\mathbf{x})$ represent the creation 
and annihilation of fermions with mass $m$ and spin state
$\sigma$, respectively at $\mathbf{x}$, $\mathcal{V}_{d}(\mathbf{x})$ 
signifies the weak random potential and $\mu$ is the chemical potential. 
We use Planck units, i.e., $\hbar=1$. In the
interaction Hamiltonian the $s$-wave fermionic interaction is defined as 
$\mathcal{V}(\mathbf{x},\mathbf{x'})=-g\delta(\mathbf{x}-\mathbf{x'})$ 
where $g$ is the bare coupling strength of fermion-fermion pairing
which we later regularize through the $s$-wave scattering length $a$. 

Recent experiments have employed different techniques to create disorder.
Most commonly used are laser speckle\cite{billy1} 
and quasi-periodic lattice\cite{roati1}.
In our calculation we consider an uncorrelated quenched disorder 
for simplicity. 
It implies that the range of the impurities should be much smaller 
than the average separation between them. 
To model it mathematically, we use the pseudo-potential as 
$\mathcal{V}_{d}(\mathbf{x})=
\sum_{i}g_{d}\delta(\mathbf{x}-\mathbf{x}_{i})$ 
where $g_{d}$ is the fermionic impurity coupling constant
(which is a function of impurity scattering length $b$),
and $\mathbf{x}_{i}$ are the static positions of the quenched disorder. 
Thus, the correlation function is
$\langle \mathcal{V}_{d}(-q)\mathcal{V}_{d}(q)\rangle=
\beta\delta_{i\omega_{m},0}\gamma$
where $q=(\mathbf{q},i\omega_{m})$. 
$\beta=1/k_BT$ is the inverse temperature,
$\omega_{m}=2\pi m/\beta$ are the 
bosonic Matsubara frequencies with $m$ an integer. 
The disorder strength can be written as $\gamma=n_{i}g_{d}^2$, where
$n_{i}$ denotes the concentration of the impurities. 
Simple algebraic manipulations further reveal that $\gamma$ is a 
function of the relative size of the impurity ($b/a$). 

The model of disorder described above can be realized in 
real experiments when a collection of light fermionic species
inside a harmonic confinement remains in the same spatial dimension 
as of the few heavy fermions in an optical lattice. The laser is
tuned in such a way that the lighter species can not see the optical 
lattice while the heavier species can not see the parabolic 
potential\cite{sademelo3}. 

We use Eq.\,(\ref{H}) and the model of disorder,
within the modified mean-field formalism through Gaussian
fluctuations. 
After carrying out the self-energy calculation using Dyson equation,
effective action can be written as\cite{orso,khan1},
\begin{eqnarray}
 \mathcal{S}_{eff}&=&\int d\mathbf{x}\int_{0}^{\beta}d\tau
\left[\frac{|\Delta(\mathbf{r})|^2}{g}
-\frac{1}{\beta}\mathrm{Tr}\ln\{-\beta \mathcal{G}^{-1}(\mathbf{r})\}\right],\label{seef}
\end{eqnarray}
where $\mathbf{r}=(\mathbf{x},\tau)$, $\Delta(\mathbf{r})$ is the 
pairing gap and $\mathcal{G}^{-1}$ is the inverse Nambu propagator.
Expansion of the inverse Nambu propagator up to the second order 
it is possible to write the effective action ($\mathcal{S}_{eff}$)
in Eq.(\ref{seef}) as a sum of bosonic action ($\mathcal{S}_{B}$) and 
fermionic action ($\mathcal{S}_{F}$). There is an 
additional term which emerges from the linear order of self-energy 
expansion ($\mathcal{G}_{0}\Sigma$). It is possible to set the linear 
order term to zero, if we consider $\mathcal{S}_{F}$ is an extremum of 
$\mathcal{S}_{eff}$, after having performed all the fermionic 
Matsubara frequency sums. The constrained condition leads to the BCS 
gap equation, which after appropriate regularization 
through $s$-wave scattering length reads,
\begin{eqnarray}
 -\frac{m}{4\pi a}=\sum_{k}\left[\frac{1}{2E_{k}}-\frac{1}{2\epsilon_{k}}\right],\label{gap}
\end{eqnarray}
where $E_{k}=\sqrt{\xi_{k}^2+\Delta^2}$, $\xi_{k}=\epsilon_{k}-\mu$, $\epsilon_{k}=k^2/(2m)$, $\mu$ is the chemical potential,
and $\Delta$ is the BCS gap function. 
Next we obtain the density equation through the thermodynamic potential ($\Omega$).
$\Omega$ can be written as a sum over fermionic ($\Omega_{F}$)
and bosonic ($\Omega_{B}$) potentials, 
$n=n_{F}+n_{B}=-\frac{\partial}{\partial\mu}(\Omega_{F}+\Omega_{B})=
-\frac{1}{\beta}\frac{\partial}{\partial\mu}(\mathcal{S}_{F}+\mathcal{S}_{B})$. 
Hence the mean-field density equation reads
\begin{eqnarray}
 n&=&\sum_{k}\Big(1-\frac{\xi_{k}}{E_{k}}\Big)-\frac{\partial\Omega_{B}}{\partial\mu},\label{density}
\end{eqnarray}
where $\Omega_{B}$ is the bosonic thermodynamic potential which 
contains the disorder contribution. 
Now Eqs.\,(\ref{gap}) and (\ref{density}) are ready to be solved 
self-consistently.

\subsection*{Fidelity Susceptibility}

Study of FS in disordered environment is a subject of interest in recent 
years, since it offers a unique intrinsic tool to detect QPT\cite{zanna,You07,Yang07,chen07}. A detailed 
description of QPT in quantum XY model in the presence of disorder has 
already been reported\cite{zanardi1,zanardi2}. Furthermore, in a recent 
analysis the role of FS in a disordered two-dimensional Hubbard model 
has been considered\cite{basu2}.
Here, we study the fidelity susceptibility within the dirty crossover 
scenario.

The ground-state fidelity $F(\lambda+\delta\lambda,\lambda)$ depends 
on both the controlling parameter $\lambda$ and its variation, 
$\delta\lambda$. The  dependence on 
$\delta\lambda$ can be eliminated by considering 
the limiting expression for the ground-state fidelity when  
$\delta\lambda$ approaches zero. For small $\delta\lambda$,  
\begin{eqnarray}
F(\lambda+\delta\lambda,\lambda)^2&=&\left[\langle\Psi (\lambda)|+
\delta\lambda \, \frac{\partial \langle\Psi (\lambda)|}{\partial\lambda}  \nonumber\right.+
\left. \frac{(\delta\lambda)^2}{2} \frac{\partial^2 \langle\Psi(\lambda)|}{\partial{\lambda}^2} 
\right] |\Psi(\lambda)\rangle\nonumber\\
 &=&1-\frac{(\delta\lambda)^2}{2}\frac{\partial\langle\Psi 
(\lambda)|}{\partial\lambda} \frac{\partial
|\Psi(\lambda)\rangle}{\partial \lambda},\label{fidelity}
\end{eqnarray} 
where the state $|\Psi(\lambda)\rangle$ is assumed to be real and 
normalized.
A sudden drop of the ground-state fidelity at the critical point will then 
correspond to a divergence of the FS\cite{You07,Yang07}:
\begin{eqnarray}
\chi(\lambda)&\equiv& - \frac{1}{V}\lim_{\delta\lambda\to 0} \frac{4 \ln F(\lambda+\delta\lambda,\lambda)}{(\delta \lambda)^2}\label{chi1}\nonumber\\
&=&\frac{1}{V}\frac{\partial\langle\Psi 
(\lambda)|}{\partial\lambda}\cdot \frac{\partial
|\Psi(\lambda)\rangle}{\partial \lambda} \label{chi2}, 
\end{eqnarray}
where $V$ is the system volume.

Since we are interested in calculating the FS across the BCS-BEC crossover,
we calculate $\chi(\lambda)$ for the BCS wave-function:
\begin{equation}
|\Psi(\lambda)\rangle=\prod_{{\bf k}}[u_{k}(\lambda)+v_{k}(\lambda)c_{{\bf k}\uparrow}^{\dagger}c_{-{\bf k}\downarrow}^{\dagger}]|0\rangle\; .\label{wavef}
\end{equation}
Here, $c_{{\bf k}\sigma}^{\dagger}$ creates a fermion in the 
single-particle state of wave-vector ${\bf k}$, spin $\sigma$ 
and energy $\epsilon_k$, $u_k$ and $v_k$ are the BCS
coherence factors, $v_k^2=1-u_k^2=(1-\xi_k/E_k)/2$.
When the BCS wave function is inserted in Eq.~(\ref{chi2}), 
one obtains
\begin{equation}
\chi(\lambda)=\int\!\!\frac{d {\bf k}}{(2\pi)^{3}} \left[\left(\frac{d u_{k}}{d\lambda}\right)^2+ \left(\frac{d v_{k}}{d\lambda}\right)^2\right],\label{chilambda}
\end{equation}
which can be written as
\begin{equation}
\chi(\lambda)=\int\!\!\frac{d {\bf k}}{(2\pi)^{3}} \frac{1}{4E_{k}^4}\left[\Delta_{k}\frac{d\mu}{d\lambda}+\xi_{k}\frac{d\Delta_{k}}{d\lambda}\right]^2
\, .\label{chiBCS}
\end{equation}
In the numerical calculation of $\chi$, we use the disorder induced 
self-consistent solution of the order parameter and the 
chemical potential where $\lambda=(k_{F}a)^{-1}$.

\section{Results and Discussion}\label{result}
We now present our FS results which are obtained through self-consistent 
calculation of Eqs.\,(\ref{gap}) and (\ref{density}) and subsequent 
numerical integration of Eq.\,(\ref{chiBCS}). In the first part we 
show the effect of disorder on FS and related statistical
analysis. Afterward we show the effects of disorder on DOS and 
discuss the possibility of a QPT. 
Let us now specify what is meant by 
weak impurity.
We know that $\gamma$ has a dimension of $k_{F}/m^2$ which leads 
to the dimensionless disorder strength
$\eta=\gamma m^2/k_{F}=(3\pi^2/4)\gamma n/ \epsilon_{F}^2$.  
This implies that the impurity density and strength 
remains much less than the particle density $n$ and Fermi energy 
$\epsilon_{F}$.
For practical purposes, one can quantify the 
impurity contribution as weak, as long as the dimensionless disorder 
strength satisfies\cite{sademelo3}  $\eta\lesssim 5$. 

\subsection*{FS with Disorder}

\begin{figure}
 \begin{center}
\includegraphics[width=6.25cm]{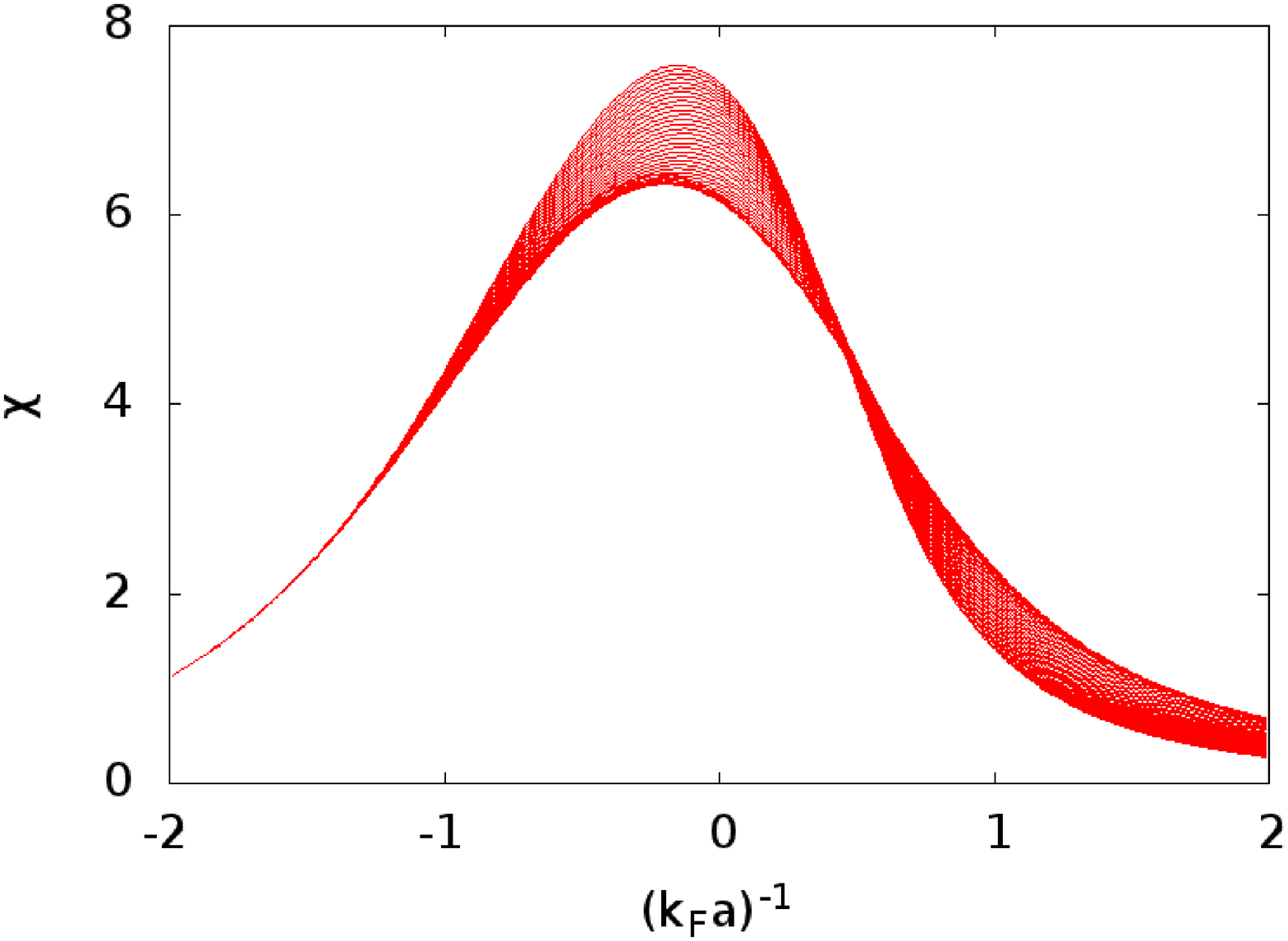}
  \includegraphics[width=6.25cm]{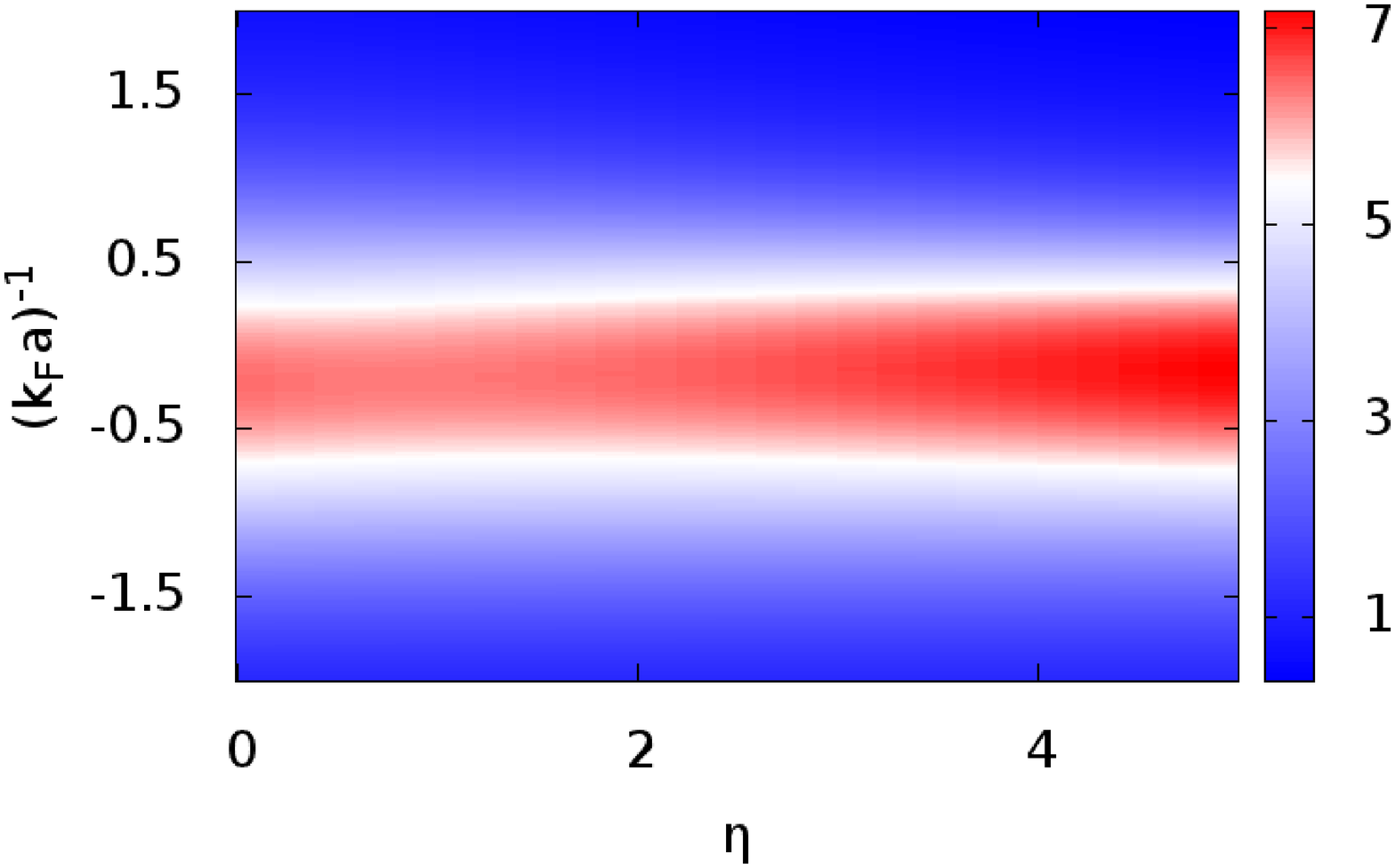}
\caption{(Color online) $\chi$ (in units of $k_{F}^{-3}$) is plotted 
for various disorder strengths along the BCS-BEC crossover. 
The left panel, each $\chi$ is separated
by an increment of $\eta=0.2$ from $\eta=0$ to $\eta=5$. The behavior 
reveals the break in symmetry for FS. In the right panel
we present the same data in density plot format for a clear understanding 
of the change in peak hight and width of the FS.}\label{FS}
 \end{center}
\end{figure}

The recent observation of Anderson localization in ultracold 
atomic gases\cite{billy1,roati1,kondov,bouyer}
has opened the possibility of studying disorder effects in 
the whole crossover region. 
This motivates us to check whether it is 
possible to predict a QPT in an interaction driven disorder induced 
ultracold Fermi system. 
Starting from the clean limit, we increase the 
impurity strength and observe a gradual change of FS from symmetric 
to asymmetric profile (Fig.(\ref{FS})). 
The left panel in Fig.\,(\ref{FS}) shows FS versus interaction for 
various disorder strengths ($\eta\in[0,5]$ with increments of 
$0.2$) and the loss of symmetry and decrease of width.
The right panel depicts a density plot between interaction, 
disorder and FS. From this plot one can realize the gradual increase in 
FS peak height and also a slow shift of the peak to the weaker to 
relatively stronger coupling. We note that in a 
study of quantum spin chains similar observations were 
also made\cite{zanardi2}. We thus 
ask whether this behavior is a precursor of an imminent phase 
transition with much stronger disorder\cite{sademelo2}. 

A measure 
of the asymmetry in FS would be a useful tool and thus we calculate 
the skewness and kurtosis of the FS. 
In nuclear physics skewness and kurtosis are established 
tools to study the phase transition from an equilibrium hadronic state
to quark-gluon plasma\cite{muller,neda}.
Furthermore, these techniques are also applicable to chiral
phase transition in quantum chromodynamics (QCD)\cite{redlich}. 
Skewness ($S$) is defined as the measure of the asymmetry of the 
distribution, and kurtosis ($\kappa$) measures the curvature 
of the distribution peak, which are defined as
\begin{eqnarray}
 S&=&\frac{\langle(x-\langle x\rangle)^3\rangle}{\langle(x-\langle x\rangle)^2\rangle^{3/2}},
\quad\kappa=\frac{\langle(x-\langle x\rangle)^4\rangle}{\langle(x-\langle x\rangle)^2\rangle^2}-3.
\end{eqnarray}
In the present context $x=(k_{F}a)^{-1}$.
The critical transition point can be identified as the point where 
$S$ and $\kappa$ change sign, since the change of sign will 
signify an abrupt non-analytic change in the orientation (for skewness) 
and height (for kurtosis) of the distribution. 
Hence, they can be used as indicators of an impending transition.

Physically, the change of sign in skewness can be viewed as change of 
direction of asymmetry from left to right or vice versa. In the
case of $\kappa$ 
it can be viewed as change of a sharp distribution to a flat one. At the phase 
transition, FS is expected to diverge
as a result of a sudden drop in fidelity\cite{khan2}. 
Here, we expect that, as we near the transition point the FS 
distribution should become increasingly sharper
resulting in an increasingly larger kurtosis.
\begin{figure}
 \begin{center}
 \includegraphics[height=6.25cm,angle=270]{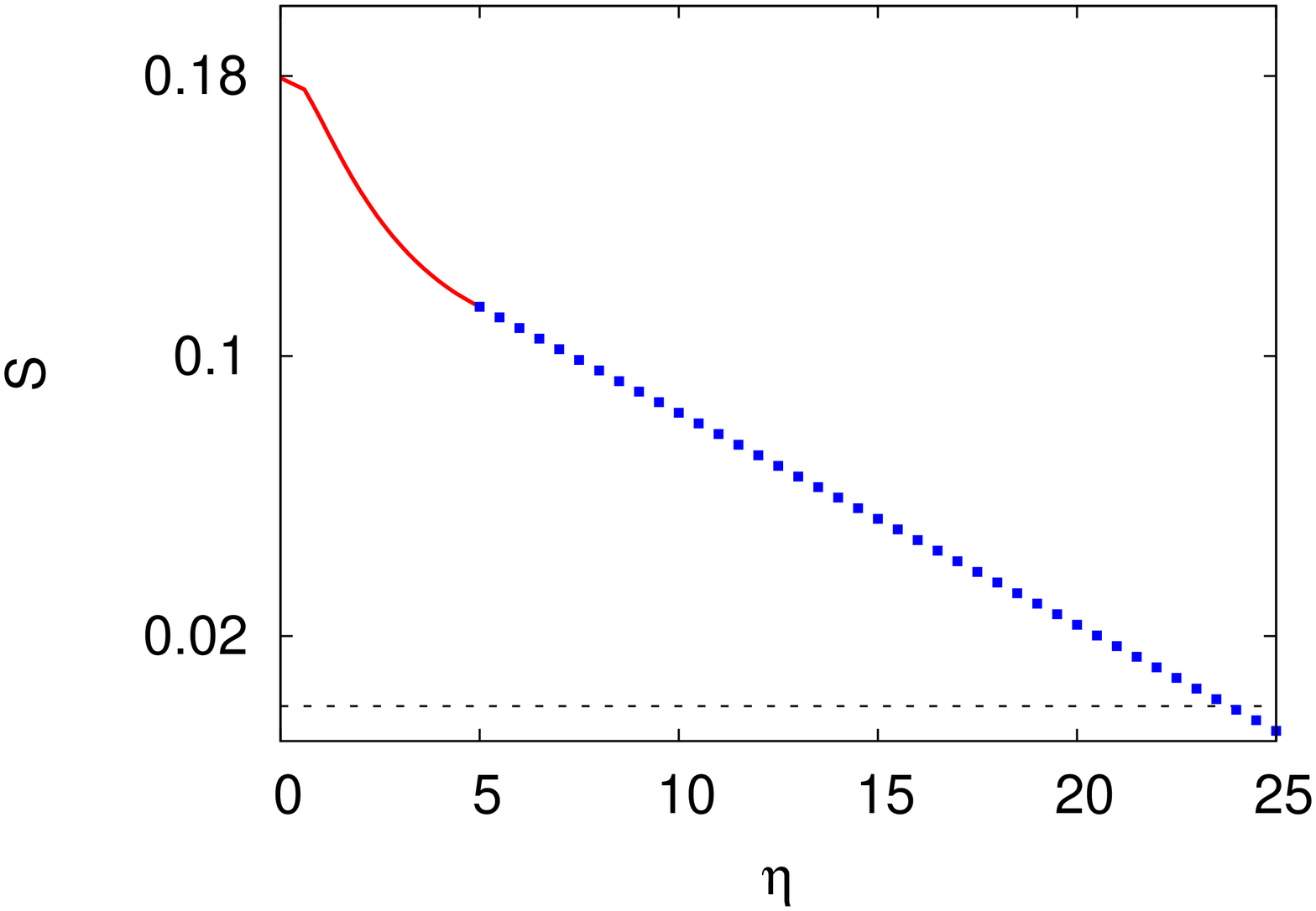}
 \includegraphics[height=6.25cm,angle=270]{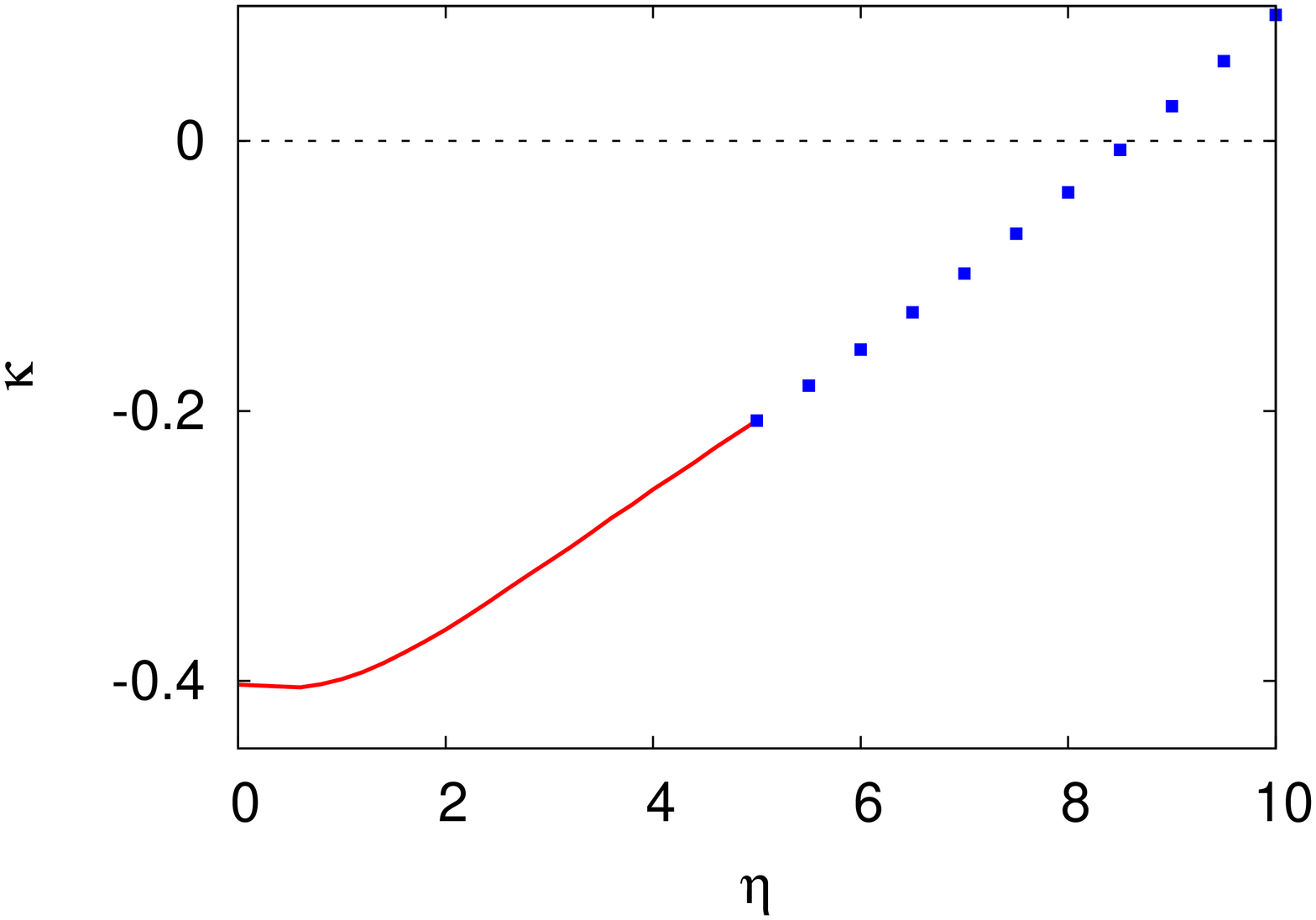}
\caption{(Color online) The left panel presents the skewness 
whereas the lower panel is the kurtosis. Our calculation of skewness and kurtosis are presented through the solid line.
The bold points depict the non-linear extrapolation and the dotted line indicates the zero axis.}\label{SK}
 \end{center}
\end{figure}

We calculate $S$ and $\kappa$ as a function 
of disorder which are depicted in Fig.(\ref{SK}).
In this figure, the left panel represents the skewness, which being 
positive implies that the FS distribution tail is longer on the BEC side. 
The right panel tells us about the kurtosis which is negative, 
and thus corresponds to a flatter peak which is usually observed in 
the phase transition\cite{neda}. The solid line depicts the calculated $S$ and $\kappa$. The bold points
are our nonlinear extrapolation of the data. 

We observe with increasing disorder the FS peak moves towards the 
relatively strong interaction regime 
and the asymmetry gradually increases which can be viewed in the 
progressive change of $S$ and $\kappa$ in Fig.(\ref{SK}).
From the Anderson theorem it is known that in
three-dimensions there exits a critical impurity strength beyond 
which the electrons will be localized.
In this study, we can not comment exactly on the localization because 
of the limitations of the modified mean-field theory. However,
it can be noticed (Fig.\,(\ref{SK})) that both $S$ and $\kappa$
approach zero where a sign change will occur (possible phase transition). 
To estimate the critical point of phase transition, we first apply 
least square fitting of the data points which indicates that 
$\eta_{c}$ is around $10-13$. As an alternative, we also employ 
nonlinear spline interpolation.
The results suggest the zero crossing for these moments may occur 
at $\eta_{c}\simeq9$ (kurtosis) and $\eta_{c}\simeq23$
(skewness).
A range of values for $\eta_{c}$ in $S$ and $\kappa$ has also
been encountered in the study of quark-gluon plasma
with the suggested interpretation of
$\eta_{c}\simeq9$ signifying the beginning of the QPT and 
$\eta_{c}\simeq23$ indicating the completion of QPT\cite{neda}.
We also note that our linearly extrapolated critical point is in 
agreement with Ref.\,\refcite{sademelo2}, where $\eta_{c}$ is 
estimated to be around $11$.
\subsection*{DOS with Disorder}

In the previous section we have considered the possibility of a QPT 
in a dirty crossover. To study the influence of disorder 
in the BCS-BEC crossover more closely, we calculate the 
DOS defined as 
\begin{eqnarray}
N(\omega)=\sum_{k}u_{k}^2\delta(\omega-E_{k})
+v_{k}^2\delta(\omega+E_{k}),\label{DOS}
\end{eqnarray}
where $u_{k}$, $v_{k}$ and $E_{k}$ are the usual BCS parameters as 
introduced earlier. 
\begin{figure}
 \begin{center}
 \includegraphics[height=7.cm]{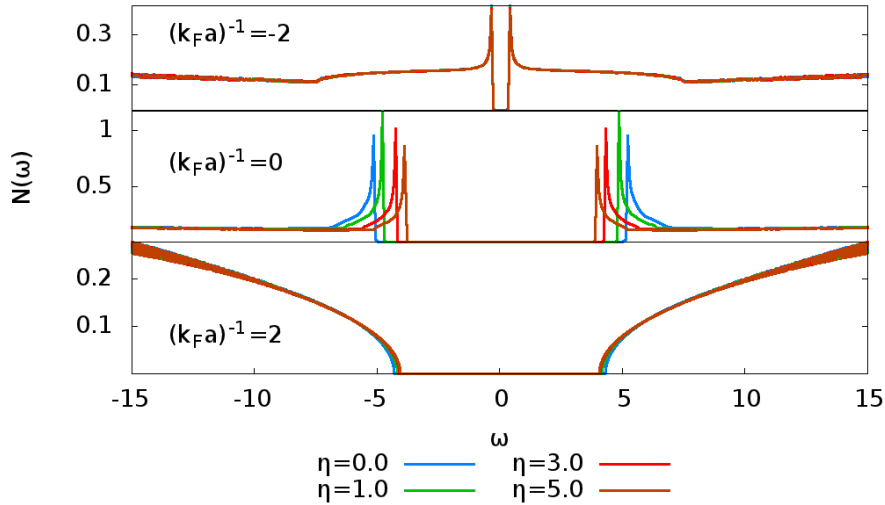}
\caption{(Color online) The density of states for different
couplings and various disorder strengths is presented. 
From top to bottom the figures describe weak coupling, unitarity 
and strong coupling regions respectively. It is noticeable that 
disorder plays a significant part only in the unitarity
leaving the other two regions almost unchanged.}\label{DOS_plot}
 \end{center}
\end{figure}

In Fig.\,(\ref{DOS_plot}) we present the behavior of DOS, 
$N(\omega)$, at different interaction regimes. 
Interestingly, we do not observe any significant difference 
in BCS and BEC limit in terms of energy gap as well as stacking 
of energy states. But at unitarity, the energy gap
gradually reduces as a function of disorder and sharp coherence 
peaks at the excitation edges are observed.
\begin{figure}
 \begin{center}
 \includegraphics[width=10.cm]{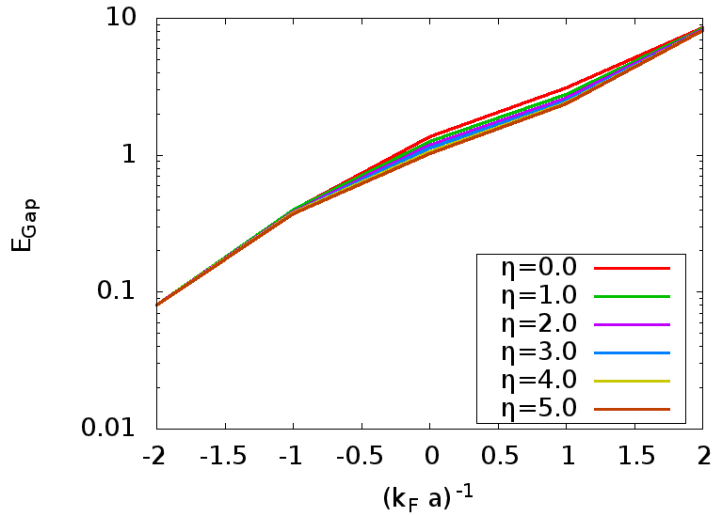}
\caption{(Color online) Variation of the energy gap (spectral gap) 
with disorder strength and interaction energy is shown. 
A distinct drop in $E_{Gap}$ can be noted in the crossover region 
due to disorder and this drop increases with increasing impurity 
strength. However the two extremes (BCS and BEC) do not show
much change.}\label{egap}
 \end{center}
\end{figure}

We examine this behavior more closely in Fig.\,(\ref{egap}). 
In the BCS-BEC crossover region 
($(k_{F}a)^{-1}\in[-1,1]$) one can observe that 
the $E_{Gap}$ is reduced, but on the BCS and BEC sides it remains unaffected by disorder.
It is interesting to note that in a recent study on coexistence 
of pseudogap and usual BCS energy gap in a harmonic trap\cite{yoji}
the reduction of energy gap was observed only at 
unitarity when the DOS was calculated at different positions 
from trap center. 
We already know that in a BCS superconductor the energy gap 
reduces at moderate disorder but starts increasing
for stronger disorder, which brings the onset of localization. 
Here the effect of disorder is more prevalent at unitarity
compared to the other regions. 
This can be taken as an indication of an early onset of QPT at unitarity 
and the emergence of a transient state (glassy phase) at 
relatively moderate disorder. 

\subsection*{Discussion}

In this work we have combined different ideas from 
condensed matter physics and quantum information theory to 
investigate the effect of disorder on the BCS-BEC
crossover. We now discus various aspects and experimental 
perspectives of the ideas presented here.

We emphasize that through a relatively simple method 
we are able to comment on the possible QPT in a disordered unitary Fermi 
gas. Apart from a detailed real space 
analysis of the inhomogeneous system (by employing a BdG analysis)
it will also be very interesting to see the effect of the impurity 
in this model by incorporating higher orders in the Dyson equation.
Such a comparative study, involving the third and fourth orders 
in the free energy expansion would be
useful in understanding the range of validity of the
perturbation theory. 

We note that Bose gas experiments were conducted with different 
disorder realizations such as speckle and incommensurate (bichromatic) 
lattice\cite{billy1,roati1}. 
To realize the proper range of localization length (smaller than 
the system size) it is necessary to design an appropriate optical 
system for speckle potential. However, the quasiperiodic lattice 
naturally provides a very short coherence length\cite{modu}. 
The uneven landscape of the impurity potential allows existence 
of spatially confined states which can result in `island' formation 
of superfluid system in an otherwise insulating domain and this feature is
expected to influence the Bose condensate more severely 
than the BCS Cooper pairs because of strong phase coherence.
The quenched disorder described here is repulsive, thus
it is non-confining. The impurity potential represents essentially 
the scattering centers which affect mostly
the phases of the wave functions of the particles that are scattered 
from them. Thus, the model does not influence any localization inherently.

We also note that the use of FS is gathering attention 
in the study of ultracold atomic systems\cite{rigol}. From the quantum 
information theory side, an experimental model 
to observe fidelity was suggested almost a decade ago\cite{ekert} 
and it is observed in photon entanglement measurements\cite{kiesel}. 
Since one can now observe the macroscopic wave functions in 
ultracold atom experiments, we expect the realization of the 
overlap of two wave functions in the near future. 

\section{Conclusion}\label{conclusion}

In conclusion, we have considered the possibility of 
QPT near unitarity through the study of fidelity susceptibility and 
density of states with quenched disorder. The disorder is included
in the mean-field formalism through Gaussian fluctuations.
The FS provides a strong indication in terms of
evolution from symmetric to asymmetric FS line-shape due to disorder 
for a QPT. A more detailed real space 
analysis with strong disorder should provide further information 
on this issue.

We have calculated the DOS for different interaction regimes.
Interestingly, the DOS shows sharp coherence peaks at the excitation 
edges and gradual lowering of the energy gap.
On the other hand, the DOS remains almost the same as that of a 
clean system in the deep BCS and BEC regimes. 
This suggests that the unitary superfluid is a better candidate 
for impending disorder driven QPT.

We note that the effect of impurities in superconducting lattice systems 
has been extensively studied. 
The emergence of ultracold Fermi gases
necessitates a detailed investigation with disorder in a 
homogeneous continuum model. 
The quenched disorder model within the BEC has been studied
through the mean-field theory and QMC 
simulations \cite{astrakharchik,huang,stringari2}. 
A systematic description across the crossover is missing.
We hope our analysis will bridge this gap and stimulate further 
investigations in the whole crossover.

\section*{Acknowledgement}
This work is supported by TUBITAK-BIDEP, TUBITAK (112T176 and
109T267) and TUBA.
AK thanks insightful discussion with S. W. Kim and D. Lippolis.
SB acknowledges financial support from DST and CSIR (SR/S2/CMP-23/2009 and 
03(1213)/12/EMR-II).

\end{document}